\documentclass[amsmath, amssymb, amsfonts, nofootinbib]{revtex4}
\usepackage{amsthm}
\usepackage{hyperref}

\theoremstyle{remark}

\theoremstyle{definition}

\begin{document}

\title{Comment on the paper\\``Does Zeeman's Fine Topology Exist?''\\at \href{http://arxiv.org/abs/1003.3703v1}{arXiv:1003.3703v1 [math-ph]}}
\date{\today}
\author{Giacomo Dossena}
\email{dossena@sissa.it}
\affiliation{SISSA - Via Bonomea 265 - 34136, Trieste - Italy}

\begin{abstract}
A constructive and straightforward proof of the existence of the Zeeman topology is provided, contradicting a fallacious claim contained in the paper ``Does Zeeman's Fine Topology Exist?'' available at \href{http://arxiv.org/abs/1003.3703v1}{arXiv:1003.3703v1}.
\end{abstract}

\maketitle
\section*{Preamble}
In the past few years two papers appeared on the arXiv claiming to prove the non-existence of the Zeeman topology. The oldest one \cite{sainz1} has been withdrawn and replaced by a new one \cite{sainz2} which shares techniques and conclusions with the oldest. A review of these papers is beyond the scope of this short communication.

Suffice to say that the existence of Zeeman's topology is made obvious by a reading of the original and beautiful paper \cite{zeeman}. For the sake of clarity a constructive proof of its existence is spelt out below, essentially rephrasing an argument of Zeeman.

\section*{Constructive proof of the existence of Zeeman's topology}
The Zeeman topology on Minkowski space $M$ is defined to be the finest topology on $M$ inducing\footnote{Given a topological space $(X,T)$ and a subset $Y\subset X$ with a topology $T_Y$, we say $(X,T)$ induces $T_Y$ on $Y$ if $(Y,T_Y)$ is a subspace of $(X,T)$. In other words, each open set in $(Y,T_Y)$ is the intersection of $Y$ with an open set in $(X,T)$.} the Euclidean topology on each timelike and spacelike affine subspace (from now on called axes). To prove its existence, let us consider the collection $C$ of all subsets of $M$ that meet each axis in an open\footnote{Open w.r.t. the Euclidean topology of the axis, of course.} set. This collection satisfies the axioms for a topology (straightforward exercise, see Lemma III.3 in \cite{dossena}) and induces the Euclidean topology on each axis by construction. Given a topology $T$ on $M$ inducing the Euclidean topology on each axis, any element of $T$ meets each axis in an open set, thereby $T$ is coarser than $C$ and the proof is complete.

\section*{Side note about Larson's theorem}
In \cite{sainz2}, Larson's theorem about maximum and minimum topological spaces is quoted. Let us review this interesting theorem (see \cite{larson}). Fix a topological property $P$. A topology $T$ on some set $X$ is called \emph{maximum P} [\emph{minimum P}] when any topology on $X$ with property $P$ is coarser [finer] than $T$. Larson's theorem states that a topology on $X$ is maximum $P$ or minimum $P$ for some topological property $P$ if and only if each bijection of $X$ onto itself is a homeomorphism. Contrary to claims in \cite{sainz2}, this theorem does not apply to Zeeman's topology since the property of inducing the Euclidean topology on each axis is not a topological property (this is obvious since its very formulation requires the concept of affine subspace, which is not topologically invariant).

\end{document}